\documentstyle[11pt]{article}

\begin{document}
\title{\bf Second order Lax pairs of nonlinear partial differential equations
with Schwarz variants}

\author{Sen-yue Lou$^{1,2,3}$\thanks{Email:
sylou@mail.sjtu.edu.cn}, Xiao-yan Tang$^{2,3}$, Qing-Ping Liu$^{1,4,3}$ and T. Fukuyama$^{5,6}$\\
 \it \footnotesize \it $^{1}$CCAST (World
Laboratory), PO Box
 8730, Beijing 100080, P. R. China\\
\it \footnotesize \it $^{2}$Physics Department of Shanghai Jiao
Tong University,
Shanghai 200030, P. R. China\\
\footnotesize \it $^{3}$Abdus Salam International Centre for
Theoretical Physics, Trieste, Italy\\
\footnotesize \it $^{4}$Beijing Graduate School, China University
of Mining and Technology,\\ \footnotesize \it Beijing 100083,
Peoples R China. Peking 100083, P. R. China\\
\footnotesize \it $^5$Department of Physics, Ritsumeikan
University, Kusatsu, Shiga 525-8577, Japan\\
\footnotesize \it $^6$Department of Physics, University of
Maryland, College Park, MD20742, U.S.A.}

\date{}

\maketitle

\begin{abstract}
In this paper, we study the possible second order Lax operators
for all the possible (1+1)-dimensional models with Schwarz
variants and some special types of high dimensional models. It is
shown that for every (1+1)-dimensional model and some special
types of high dimensional models which possess Schwarz variants
may have a second order Lax pair. The explicit Lax pairs for
(1+1)-dimensional Korteweg de Vries equation, Harry Dym equation,
Boussinesq equation, Caudry-Dodd-Gibbon-Sawada-Kortera equation,
Kaup-Kupershmidt
 equation, Riccati equation, (2+1)-dimensional breaking soliton equation and a generalized (2+1)-dimensional
fifth order equation are given.
\end{abstract}
\vskip.1in

\section{Introduction}

In the study of a nonlinear mathematical physics system, if one can find that
the nonlinear system can be considered as a consistent condition
of a pair of a linear problem, then some types of special
important exact solutions of the nonlinear system can be solved by
means of the pair of linear problem. The pair of linear system is
called as Lax pair of the original nonlinear system and the
nonlinear system is called as Lax integrable or IST (inverse
scattering transformation) integrable. Usually, a Lax integrable
model may have also many other interesting properties, like the
existing of infinitely many conservation laws and infinitely many
symmetries, multi-soliton solutions, bilinear form, Schwarz
variants, multi-Hamiltonian structures, Painlev\'e property etc.
In the recent studies, we found that the existence of the Schwarz
variants may plays an important role. Actually, in our knowledge,
almost all the known IST integrable (1+1)- and (2+1)-dimensional
models possess Schwarz invariant forms which is invariant under
the M\"obious transformation (conformal invariance)$\cite{Weiss, Nucci}$.
The conformal invariance of the well known Schwarz Korteweg
de-Vries (SKdV) equation is related to the infinitely many
symmetries of the usual KdV equation$\cite{SchKdV}$. The conformal
invariant related flow equation of the SKdV is linked with some
types of (1+1)-dimensional and (2+1)-dimensional sinh-Gordon (ShG)
equations and Mikhailov-Dodd-Bullough (MDB)
equations$\cite{Riccati}$. It is also known that by means of the
Schwarz forms of many known integrable models, one can find also
many other integrable properties like the B\"acklund
transformations and Lax pairs$\cite{Weiss}$. In $\cite{Lou}$, one
of the present authors (Lou) proposed that starting from a
conformal invariant form may be one of the best way to find
integrable models especially in high dimensions. Some types of
quite general Schwarz equations are proved to be Painlev\'e
integrable. In $\cite{PRL}$, Conte's conformal invariant
Painlev\'e analysis$\cite{Conte}$ is extended to obtain high
dimensional Painlev\'e integrable Schwarz equations
systematically. And some types of physically important high
dimensional nonintegrable models can be solved approximately via
some high dimensional Painlev\'e integrable Schwarz
equations$\cite{Ruan}$.

Now an important question is what kind of Schwarz equations are
related to some Lax integrable models? To answer this question
generally in arbitrary dimensions is still quite difficult. So in
this paper we restrict our main interests to discuss in
(1+1)-dimensional models.

In the next section, we prove that for any (1+1)-dimensional
Schwarz model there may be a second order Lax pair linked with it.
In section 3, we list various concrete physically significant
examples. In section 4, we discuss some special extensions in
higher dimensions. The last section is a short summary and
discussion.

\section{A second order (1+1)-dimensional Lax pair Linked with an
arbitrary Schwarz form}

In (1+1)-dimensions, the only known independent conformal
invariants are
\begin{eqnarray}
&&p_1\equiv \frac{\phi_t}{\phi_x},\\
&&p_2\equiv \{\phi; \ x\}\equiv
\frac{\phi_{xxx}}{\phi_x}-\frac32\frac{\phi_{xx}^2}{\phi_x^2},\\
&&p_3\equiv \{\phi;\ t\}\equiv
\frac{\phi_{ttt}}{\phi_t}-\frac32\frac{\phi_{tt}^2}{\phi_t^2},\\
&&p_4\equiv \{\phi;x;t\}\equiv
\frac{\phi_{xxt}}{\phi_t}-\frac{\phi_{xx}\phi_{xt}}{\phi_x\phi_t}-\frac12\frac{\phi_{xt}^2}{\phi_t^2},\\
&&p_5\equiv \{\phi;t;x\}\equiv
\frac{\phi_{xtt}}{\phi_x}-\frac{\phi_{tt}\phi_{xt}}{\phi_x\phi_t}-\frac12\frac{\phi_{xt}^2}{\phi_x^2},
\end{eqnarray}
where $\phi$ is a function of $\{x,\ t\}$, the subscripts are
usual derivatives while $\{\phi;\ x\}$ is the Schwarz derivative.
As in $\cite{Lou,PRL,Ruan}$, we say a quantity is a conformal
invariant if it is invariant under the M\"obious transformation
\begin{eqnarray}
\phi\rightarrow \frac{a\phi+b}{c\phi+d},\ ad\neq bc.
\end{eqnarray}

From (1)--(5), we know that the general (1+1)-dimensional
conformal invariant Schwarz equation has the form
\begin{eqnarray}
F(x,t,p_i,p_{ix},p_{it},p_{ixx},... (i=1,...,5) )\equiv F(p_1,\
p_2,\ p_3,\ p_4,\ p_5)=0,
\end{eqnarray}
where $F$ may be an arbitrary function of $x,\ t,\ p_i$ and any
order of derivatives and even integrations of $p_i$ with respect
to $x$ and $t$. According to the idea of $\cite{Lou}$, (7) (or many of (7)) may be
integrable. If $F$ of (7) is a polynomial function of $p_i$ and
the derivatives of $p_i$, then one may prove its Painlev\'e
integrability by using the method of $\cite{Lou,PRL}$. However,
for general function $F$ in (7), it is difficult to prove its
Painlev\'e integrability. Fortunately, we can find its relevant
variant forms with Lax pair. To realize this idea, we
consider the following second order Lax pair:
\begin{eqnarray}
&&\psi_{xx}=u\psi_x+v\psi,\\
&&\psi_t=u_1\psi_x+v_1\psi,
\end{eqnarray}
where $u,\ u_1,\ v,$ and $v_1$ are undetermined functions. To link
the Lax pair (8) and (9) with the Schwarz equation (7), we suppose
that $\psi_1$ and $\psi_2$ are two solutions of (8) and (9), and
$\phi$ of (7) is linked to $\psi_1$ and $\psi_2$ by
\begin{eqnarray}
\phi=\frac{\psi_1}{\psi_2}.
\end{eqnarray}
Now by substituting (10) with (8) and (9) into (7) directly, we
know that if the functions $u,\ v$ and $u_1$ are linked by
\begin{eqnarray}
F(P_1,\ P_2,\ P_3,\ P_4,\ P_5)=0
\end{eqnarray}
with
\begin{eqnarray}
&&P_1=u_1,\ P_2=u_x-\frac12u^2-2v,\\
&&P_3=P_2u_1^2+u_1u_{1xx}-\frac12u_{1x}^2
+u_{1xt}+u_1^{-1}(u_{1tt}-u_{1x}u_{1t})-\frac32u_1^{-2}u_{1t}^2,\\
&&P_4=P_2+u_{1xx}u_1^{-1}-\frac12u_1^{-2}u_{1x}^2,\\
&&P_5=u_1^2P_4+u_{1xt}-u_1^{-1}u_{1t}u_{1x},
\end{eqnarray}
then the corresponding nonlinear equation system for the fields
$u,\ v,\ u_1$ and $v_1$ has a Lax pair (8) and (9) while the
fields $u,\ v,\ u_1$ and $v_1$ are linked to the field $\phi$ by
the non-auto-B\"acklund transformation
\begin{eqnarray}
p_i=P_i, (i=1,\ 2,\ ...,\ 5).
\end{eqnarray}
Finally to find the evolution equation system is a straightforward
work by calculating the compatibility condition of (8) and (9),
\begin{eqnarray}
\psi_{xxt}=\psi_{txx}.
\end{eqnarray}
The result reads
\begin{eqnarray}
v_t=v_{1xx}+2vu_{1x}+u_1v_x-uv_{1x},
\end{eqnarray}
and
\begin{eqnarray}
u_t=u_{1xx}+2v_{1x}+(uu_1)_x
\end{eqnarray}
in addition to the constraint (11). In Eqs. (11), (18) and (19),
one of four functions $u,\ u_1,\ v,$ and $v_1$ remains still free.
For simplicity, one can simply take
\begin{eqnarray}
u=0,\ v_1=-\frac12u_{1x}.
\end{eqnarray}
Under the simplification (20), the final evolution equation
related to the Schwarz form (7) read
\begin{eqnarray}
v_t=-\frac12u_{1xxx}+2vu_{1x}+u_1v_x
\end{eqnarray}
with (11) for $u=0$ while the Lax pair is simplified to
\begin{eqnarray}
&&L\psi\equiv (\partial_x^2-v)\psi=0,\\
&&\psi_t=M\psi\equiv (u_1\partial_x-\frac12u_{1x})\psi.
\end{eqnarray}
It should be emphasized again that the Lax operator given in (22)
is only a second order operator.

To see the results more concretely, we discuss some special
physically significant models in the following section.

\section{Special examples}

\qquad From the suitable selections of $F\equiv F(p_1,\ p_2,\
p_3,\ p_4,\ p_5)$ of (7), we may obtain various interesting examples according
to the general theory of the last section.

\leftline{\bf Example 1. \rm KdV equation.}

For the KdV equation, its Schwarz variant has the simple
form
\begin{eqnarray}
F_{KdV}(p_i)=p_1+p_2=0.
\end{eqnarray}
According to the formula (11) with $u=0$, we
know that the relation between the functions $v$ and $u_1$ is
simply given by
\begin{eqnarray}
v=\frac12 u_1.
\end{eqnarray}
Substituting (25) into (22) and (23), we re-obtain the well known
Lax pair
\begin{eqnarray}
&& \psi_{xx}-\frac12u_1\psi=0,\\
&& \psi_t=u_1\psi_x-\frac12u_{1x}\psi
\end{eqnarray}
for the KdV equation
\begin{eqnarray}
u_{1t}=3u_1u_{1x}-u_{1xxx}.
\end{eqnarray}

\leftline{\bf Example 2. \rm Harry-Dym (HD) equation.}

For the HD equation, the Schwarz form reads
\begin{eqnarray}
F_{HD}(p_i)=p_1^2-\frac2{p_2}=0
\end{eqnarray}
which leads to the relation between the functions $v$ and $u_1$
by
\begin{eqnarray}
v=\frac1{u_1^2}.
\end{eqnarray}
From (22), (23) and (30), one can obtain the known Lax pair
\begin{eqnarray}
&& \psi_{xx}-\frac1{u_1^2}\psi=0,\\
&& \psi_t=u_1\psi_x-\frac12u_{1x}\psi
\end{eqnarray}
for the HD equation
\begin{eqnarray}
u_{1t}=\frac14u_1^3u_{1xxx}.
\end{eqnarray}

\leftline{\bf Example 3. \rm Modified Boussinesq Equation and
Boussinesq equation}

For the modified Boussinesq (MBQ) equation (and the Boussinesq
equation), the Schwarz form has the form
\begin{eqnarray}
F_{MBQ}(p_i)=p_{2x}+3p_1p_{1x}+3p_{1t}=0.
\end{eqnarray}
Using (34) and (11), we have
\begin{eqnarray}
v=\frac34u_1^2+\frac32\int u_{1t}{\rm dx}.
\end{eqnarray}
Substituting (35) into (22) and (23) we get a Lax pair
\begin{eqnarray}
&& \psi_{xx}-\left(\frac34u_1^2+\frac32\int u_{1t}{\rm dx}\right)\psi=0,\\
&& \psi_t=u_1\psi_x-\frac12u_{1x}\psi.
\end{eqnarray}
The related compatibility condition of (36) and (37) reads
\begin{eqnarray}
3u_1^2u_{1x}+3u_{1x}\int u_{1t}{\rm dx}-\frac12 u_{1xxx}-\frac32 \int
u_{1tt} {\rm dx} =0.
\end{eqnarray}
Eq. (38) is called as the modified Boussinesq equation because it
is linked with the known Boussinesq equation
\begin{eqnarray}
u_{tt}+\left(3u^2+\frac13u_{xx}\right)_{xx}=0
\end{eqnarray}
by the Miura transformation
\begin{eqnarray}
u=\frac13(\pm u_{1x}-u_1^2-\int u_{1t} {\rm dx}).
\end{eqnarray}

\leftline{\bf Example 4. \rm Generalized fifth order KdV (FOKdV)
equation.}

The generalized fifth order Schwartz KdV equation has the form
\begin{eqnarray}
F_{FOKdV}(p_i)=p_1-a_1p_{2xx}-a_2p_2^2=0,
\end{eqnarray}
where $a_1$ and $a_2$ are arbitrary constants. Using (41) and
(11), we have
\begin{eqnarray}
u_1=-2a_1v_{xx}+4a_2v^2.
\end{eqnarray}
Substituting (42) into (22) and (23) we get
\begin{eqnarray}
&& \psi_{xx}-v\psi=0,\\
&& \psi_t=(a_1v_{xxx}-4a_2vv_x)\psi-2(a_1v_{xx}-2a_2v^2)\psi_x.
\end{eqnarray}
The related compatibility condition of (43) and (44) is the
generalized FOKdV equation
\begin{eqnarray}
v_t-a_1v_{xxxxx}+4(a_1+a_2)vv_{xxx}+2(a_1+6a_2)v_xv_{xx}-20a_2v^2v_x=0.
\end{eqnarray}
Some well known fifth order integrable patial differential
equations are just the special cases of (45). The usual FOKdV
equation is related to (45) for
\begin{eqnarray}
a_1=1,\ a_2=\frac32.
\end{eqnarray}
The Caudry-Dodd-Gibbon-Sawada-Kortera equation is related to (45)
for
\begin{eqnarray}
a_1=1,\ a_2=\frac14
\end{eqnarray}
while the parameters $a_1$ and $a_2$ for the Kaup-Kupershmidt equation
read
\begin{eqnarray}
a_1=1,\ a_2=4.
\end{eqnarray}

\leftline{\bf Example 5. \rm Generalized seventh order KdV (SOKdV)
equation.}

The generalized seventh order Schwartz KdV equation has the form
\begin{eqnarray}
F_{SOKdV}(p_i)=p_1-p_{2xxxx}-\alpha p_2p_{2xx} -\beta
p_{2x}^2-\lambda p_2^3 =0,
\end{eqnarray}
where $\alpha$ $\beta$ and $\lambda$ are arbitrary constants. Using (49) and
(11), we have
\begin{eqnarray}
u_1=-2v_{xxxx}+4\alpha vv_{xx} +4\beta v_x^2-8\lambda v^3.
\end{eqnarray}
Substituting (50) into (22) and (23) we get
\begin{eqnarray}
&& \psi_{xx}-v\psi=0,\\
&& \psi_t=(v_{xxxxx}-2(\alpha+2\beta) v_xv_{xx} -2\alpha vv_{xxx}
+12\lambda v^2v_x)\psi\nonumber\\
&&\qquad  +(-2v_{xxxx}+4\alpha vv_{xx}+4\beta v_x^2 -8\lambda
v^3)\psi_x.
\end{eqnarray}
The related compatibility condition of (50) and (51) is the
generalized SOKdV equation
\begin{eqnarray}
&&v_t-v_{xxxxxxx}+2(\alpha+2)vv_{xxxxx}+2(1+2\beta+3\alpha)v_xv_{xxxx}
-(16\beta +12\alpha +72\lambda)vv_xv_{xx}
\nonumber\\
&&\qquad +(8\alpha v_{xx}-8\alpha v^2+4\beta v_{xx}-12\lambda
v^2)v_{xxx} -(24\lambda+4\beta) v_x^3+56\lambda v^3v_x=0.
\end{eqnarray}
The usual SOKdV equation is related to (53) for
\begin{eqnarray}
\alpha=5,\ \beta=\frac52,\ \lambda=\frac52.
\end{eqnarray}
The seventh order CDGSK equation corresponds to
\begin{eqnarray}
\alpha=12,\ \beta=6,\ \lambda=\frac{32}3.
\end{eqnarray}
The parameters of the seventh order KK equation can be read from
\begin{eqnarray}
\alpha=\frac32,\ \beta=\frac34,\ \lambda=\frac16.
\end{eqnarray}

\leftline{\bf Example 6. \rm Riccati equation (RE) }

If the Schwarz form (7) is simply taken as
\begin{eqnarray}
F_{SKdV}(p_i)\equiv p_5=0,
\end{eqnarray}
then we have
\begin{eqnarray}
v=\frac12u_1^{-1}u_{1xx}+\frac14u_{1}^{-2}(2u_{1xt}-u_{1x}^2)-\frac12u_1^{-3}u_{1t}u_{1x}.
\end{eqnarray}
The evolution equation of $u_1$ reads
\begin{eqnarray}
3u_1u_{1t}u_{1xt}-u_1^2u_{1xtt}+u_1u_{1x}u_{1tt}-3u_{1x}u_{1t}^2=0,
\end{eqnarray}
while the related Lax pair reads
\begin{eqnarray}
&& \psi_{xx}-(\frac12u_1^{-1}u_{1xx}+\frac14u_{1}^{-2}(2u_{1xt}-u_{1x}^2)-\frac12u_1^{-3}u_{1t}u_{1x})\psi=0,\\
&& \psi_t=(-\frac12u_{1x}+\lambda_1)\psi+u_1\psi_x.
\end{eqnarray}
Actually (59) is equivalent to a trivial linearizable Riccati equation
\begin{eqnarray}
w_t=w^2+f_1(x)
\end{eqnarray}
under the transformation
\begin{eqnarray}
u_1=\exp \left(2\int w{\rm dt}\right)
\end{eqnarray}
where $f_1(x)$ is an arbitrary function of $x$. It is worth to
mention again that the well known (1+1)-dimensional ShG model and
MDB model are just the non-invertable Miura type deformation of
the Riccati equation$\cite{Riccati}$.

\section{Special extensions in higher dimensions}

\qquad From section 2, we know that the key procedure to find Lax
pair from the general conformal invariant form (7) is to find a
suitable Lax form ansatz (like (8) and (9)) and a suitable
relation ansatz (like (10) between the field of the Schwarz form
and the spectral function such that all the conformal invariants
($p_i$) becomes spectral function independent variables ($P_i$).

To extend this idea to high dimension is quite not easy. We hope
to solve this problem in future studies. In this section we give
out some special extensions in high dimensions with the same Lax
pair forms of (8) and (9).

If all the fields are functions of not only $\{x,\ t\}$, but also
$\{y,\ z,\ ...,\}$ then all the formal theory is still valid if
the independent conformal invariants of (11) is still restricted
as $p_i,\ i=1,...,5$ while the function $F$ of (11) may also
include some derivatives and integrations of $p_i$ with respect to
other space variables $y,\ z,\ ...$ etc. Here we list only two
special examples:

\leftline{\bf Example 7. \rm (2+1)-dimensional KdV type breaking soliton equation}

The concept of breaking soliton equations is firstly developed by $\cite{BS1}$
and $\cite{BS2}$ by extending the usual constant spectral problem to non-constant spectral preblem.
Various interesting properties of the breaking soliton equations have been revealed by many authors.
For instance, infintely many symmetries of some breaking soliton equations are
given in
 $\cite{LiYS, LOUSym}$. In $\cite{strong}$, it is pointed out that
 every (1+1)-dimensional
 integrable model can be extended to some higher dimensional
 breaking soliton equations with help of its strong symmetries. Yu and Toda $\cite{Yu}$ had given
out the Schwarz form of the (2+1)-dimensional KdV type
breaking soliton equation
\begin{eqnarray}
F_{2dSKdV}\equiv p_1+\int p_{2y} {\rm dx}=0.
\end{eqnarray}
From (11) and (64), we have
\begin{eqnarray}
u_1=2\int v_y  {\rm dx}.
\end{eqnarray}
Substituting (65) into (22) and (23), we obtain a Lax pair
\begin{eqnarray}
&& \psi_{xx}=v\psi,\\
&& \psi_t=2\int v_y  {\rm dx}\psi_x -(v_y-\lambda_1)\psi.
\end{eqnarray}
for the (2+1)-dimensional KdV type breaking soliton equation
\begin{eqnarray}
v_t=-v_{xxy}+4vv_y+2v_x\int v_y  {\rm dx}\equiv \Phi v_y,
\end{eqnarray}
where $\Phi$ is just the strong symmetry of the (1+1)-dimensional KdV
equation.

\leftline{\bf Example 8. \rm (2+1)-dimensional fifth order equation}

If we make the replacement
\begin{eqnarray}
p_2\rightarrow \int p_{2y} {\rm dx}
\end{eqnarray}
for some of $p_2$ in all the examples of the Last section, then we
can obtain some special types of their (2+1)-dimensional
extensions. Example 8 is just obtained from the (1+1)-dimensional KdV
equation by using the replacement (69).

A generalization of the
fifth order
 Schwarz equation (41) reads ($b_1+b_2=a_1,\ c_1+c_2+c_3=a_2$),
\begin{eqnarray}
 p_1=b_1p_{2xy}+b_2p_{xx}+c_1\left(\int p_{2y} {\rm dx}
\right)^2+c_2p_2\int p_{2y} {\rm dx}+c_3 p_{2}^2.
\end{eqnarray}
From (11) and (70) we know
\begin{eqnarray}
u_1=-2b_1v_{xy}-2b_2v_{xx}+4c_1\left(\int v_{y} {\rm dx}
\right)^2+4c_2v\left(\int v_{y} {\rm dx} \right)+4c_3v^2
\end{eqnarray}
and the related Lax pair becomes
\begin{eqnarray}
&& \psi_{xx}=v\psi,\\
&& \psi_t=(4c_1(\int v_{y} {\rm dx} )^2+4c_2v(\int v_{y} {\rm dx}
)+4c_3v^2-2b_1v_{xy}-2b_2v_{xx})\psi_x\nonumber\\
&&\qquad +(\lambda_1-2v_x(2c_3v+c_2\int v_y{\rm dx})
-2v_y(2c_1\int v_{y} {\rm dx}+c_2v)+b_1v_{xxy}+b_2v_{xxx})\psi.
\end{eqnarray}
while the corresponding evolution for the field $v$ is
\begin{eqnarray}
&&v_t=b_1v_{xxxxy}+b_2v_{xxxxx}
+2v_2(4c_2v^2-6c_1v_{xy}-3c_2v_{xx}+8c_1v\int v_y {\rm
dx})\nonumber\\
&& \qquad-2(c_2v+2b_1v+2c_1\int v_y {\rm
dx})v_{xxy}-2(2c_3v+2b_2v+c_2\int v_y {\rm dx})v_{xxx}\nonumber \\
&&\qquad +2(10c_3v^2+2c_1(\int v_y {\rm dx})^26c_2v\int v_y {\rm
dx}-(b_2+6c_3)v_{xx}-(b_1+3c_2)v_{xy}).
\end{eqnarray}
It is obvious that when $y=x$ and/or $v_y=0$, (2+1)-dimensional
fifth order equation (74) will be reduced back to (1+1)-dimensional
FOKdV equation (45).

\section{On spectral parameters}

In the last two sections, we have omitted the spectral
parameter(s). In order to add the possible spectral parameter(s)
to the Lax pairs, we may use the symmetry transformations of the
original nonlinear models. In some cases, to find a symmetry
transformation such that a nontrivial parameter can be included in
the Lax pair (8) and (9) is quite easy. For instance, it is well
known that the KdV equation (28) is invariant under the Galileo
transformation
\begin{eqnarray}
u_1\rightarrow u_1(x+3\lambda t,\ t)+\lambda \equiv u_1(x',\
t)+\lambda .
\end{eqnarray}
Substituting (75) into (26) and (27) yields the usual Lax pair of
the KdV equation with spectral parameter $\lambda$:
\begin{eqnarray}
&& \psi_{xx}-\frac12(u_1+\lambda)\psi=0,\\
&& \psi_t=(u_1-2\lambda)\psi_x-\frac12u_{1x}\psi,
\end{eqnarray}
where $x'$ has been rewritten as $x$.

However, for some other models to add the parameters to (8) and
(9) is quite difficult. In other words, the spectral parameters
may be included in (8) and (9) in very complicated way(s). For
instance, for the CDGSK equation ((45) with (47)), we failed to
include a nontrivial spectral parameter by using its point Lie
symmetries. Nevertheless, if we use the higher order symmetries
and/or nonlocal symmetries of the model, we can include some
nontrivial parameters in (43) and (44) with (47). For instance,
for the CDGSK equation, if $\psi_1$ is a special solution of (43)
and (44) with (47), one can prove that
\begin{eqnarray}
u'=u-6\frac{\lambda(\lambda\psi_1^2-\lambda \psi_{1x}
p-6\psi_{1x})}{(\lambda p+6)^2}
\end{eqnarray}
with
\begin{eqnarray}
p_x=\psi_1
\end{eqnarray}
is also a solution of the CDGSK equation. By substituting (78) into (43) and (44),
we obtain a second order Lax pair ($P=6+\lambda p$)
\begin{eqnarray}
&& \psi_{xx}=-\left(u-6\frac{\lambda(\lambda\psi_1^2-\lambda
\psi_{1x} p-6\psi_{1x})}{(\lambda p+6)^2}\right)\psi,\\
&&\psi_t=\left(\frac{6\lambda(u_x\psi_1)_x}{P}-12\frac{\psi_1\lambda^2(3u\psi_{1x}
+2u_x\psi_1)}{P^2}+72\lambda^3\psi_1\frac{u\psi_1^2-\psi_{1x}^2}{P^3}\right.\nonumber\\
&&\left.\qquad
-u_{xxx}-ww_x-36\lambda^4\psi_1^3\frac{2\lambda\psi_1^2-5\psi_{1x}P}{P^5}
\right)\psi\nonumber\\
&&\qquad +\left(36\lambda^2u\frac{\psi_1^2}{P^2}+2u_{xx}+u^2
-12\lambda\psi_{1x}\frac{u_x}P-36\lambda^4\frac{\psi_1^4}{P^4}
+72\lambda^3\psi_1^2\frac{\psi_{1x}}{P^3} \right)\psi_x
\end{eqnarray}
for the CDGSK model with $\psi_1$ being a solution of (43) and (44).

\section{Summary and discussions}

In summary, every (1+1)-dimensional equation which has a Shwarz variant
may possess a second order Lax pair. In this paper, we prove the
conclusion when the Schwarz form is an arbitrary function of five conformal invariants and their
any order derivatives and integrations.

Usually, the Lax operators for various integrable models (except for the KdV hierarchy)
are taken as higher order operator. Though the order of the Lax pair operators for some models have
been lower down, the spectral parameter have been disappeared. In order to recover some types of nontrivial
spectral parameters, we have to use the symmetries of the original nonlinear equations and the spectral parameter(s)
would be appeared in the second order Lax operator in some complicated ways. How to obtain some other integrabilities
from the Lax pairs listed here for general or special models is worthy of study further though the spectral parameters
have not yet been included in explicitly. One may obtain many interesting properties of some special models from the
Lax pairs without spectral parameters$^{\cite{sym, Fukuyama}}$. For instance, infinitely many nonlocal symmetries of
the KdV equation, HD equation, CDGSK equation and the KK equation can be obtained from the spectral parameter independent
Lax pairs$^{\cite{sym}}$.

The conclusion for the general (1+1)-dimensional Schwarz equations can also be extended to
some special types of (2+1)-dimensional models, like the breaking soliton equations. However, how to extend the
method and the conclusions to general (2+1)-dimensions or even in higher dimensions is still open.

\vskip.2in

The work was supported by the Outstanding Youth
Foundation and the National Natural Science Foundation of China
(Grant. No. 19925522), the Research Fund for the Doctoral Program
of Higher Education of China (Grant. No. 2000024832) and the
Natural Science Foundation of Zhejiang Province, China. The author
is in debt to thanks the helpful discussions with the professor
G-x Huang and the Drs. S-l Zhang, C-l Chen
and B. Wu.

\end{document}